\documentclass[conference, a4paper]{IEEEtran}
\IEEEoverridecommandlockouts
\usepackage{cite}
\usepackage[english]{babel}
\usepackage{framed}
\usepackage[normalem]{ulem}
\usepackage{amsthm}
\usepackage{amssymb}
\usepackage[scr=dutchcal,cal=euler]{mathalfa}
\usepackage{enumerate}
\usepackage[utf8]{inputenc}
\usepackage{hyperref}
\hypersetup{
    colorlinks=true,
    linkcolor=blue,
    citecolor=blue,
    filecolor=magenta,      
    urlcolor=cyan,
}
\usepackage{amsmath,amssymb,amsfonts}
\usepackage{algorithm,algorithmic}
\usepackage{graphicx}
\usepackage{mathtools}
\usepackage{gensymb}
\usepackage{xcolor}
\usepackage{subcaption}
\usepackage{tabularx,booktabs}
\usepackage{multirow,multicol}
\usepackage{stfloats}

\usepackage{fancyhdr}
\usepackage{caption}
\usepackage{tikz}
\usetikzlibrary{arrows}
\def\BibTeX{{\rm B\kern-.05em{\sc i\kern-.025em b}\kern-.08em
    T\kern-.1667em\lower.7ex\hbox{E}\kern-.125emX}}

\tikzstyle{int}=[draw, fill=blue!20, minimum size=3em]
\tikzstyle{sum}=[draw, circle, minimum size=0.5em]
\tikzstyle{init} = [pin edge={to-,thin,black}]

\makeatletter
\renewcommand*\env@matrix[1][*\c@MaxMatrixCols c]{%
  \hskip -\arraycolsep
  \let\@ifnextchar\new@ifnextchar
  \array{#1}}
\makeatother

\begin{document}

\title{
Adaptive Covariance Kalman Filtering and Nonlinear Decoupling Control via Feedback Linearization for a Three-Tank Process
}

\author{Bambang L. Widjiantoro$^{1,*}$, Katherin Indriawati$^{1}$, Moh Kamalul Wafi$^{1}$
\thanks{$^{1}$Widjiantoro, Indriawati, and Wafi is with Laboratory of Embedded and Cyber-Physical Systems, Engineering Physics Dept.,
        Institut Teknologi Sepuluh Nopember, 60111, Indonesia,
        {\tt\small b.lelono@its.ac.id}}%
}
\maketitle
\thispagestyle{fancy}

\begin{abstract}
Hydraulic three-tank systems are widely used in water treatment and liquid storage applications, where accurate level regulation is essential for safe and efficient operation. This paper investigates linear and nonlinear control strategies for reference tracking in a three-tank process. A linear state-feedback controller with integral action is first designed based on a linearized model, followed by a nonlinear decoupling controller using feedback linearization. In addition, an adaptive covariance Kalman filter (AKF) is employed for state estimation by dynamically updating the process-noise covariance matrix. Numerical simulations demonstrate that both control approaches achieve satisfactory reference tracking, while the proposed AKF provides accurate state estimation and effectively captures the nonlinear system behavior. The results highlight the effectiveness of combining nonlinear control and adaptive state estimation for hydraulic process systems.
\end{abstract}
\allowdisplaybreaks


\section{Introduction}
In the current industrial systems functioning either for liquid storage or flow treatment, hydraulic plant has becoming a big preference. This is due to the fact that chemical reactions would interfere those processes working around the desired performance so that maintaining the levels is a key requirement to reach certain goals. To deal with, various research has been proposed implementing a hydraulic mini-plant with divergent scenarios created in \cite{R1,R2} which is applied in this paper too, from the trivial linear under some desired points to the advanced uncertain non-linear systems approaching the true system as written in \cite{R3,R4} in general. The famed hydraulic system in \cite{R1} has been extensively studied among researcher in the field of optimal control and deterministic-based estimation. The decoupling constant observer is proposed in \cite{R5} to deal with FDI without fault estimation. As for the non-linear, the bi-linear approach Luenberger observer are studied by \cite{R6} and \cite{R7} regarding the leakage detection and the FDI using the cooperation of the likelihood ratio and the updated innovation over time respectively. Beyond that, the robust design of sliding mode scenario to estimate the fault with some occurrence of noises is written in \cite{R8}. In lie of advanced non-linear model, \cite{R9,R10} have developed array of decoupled estimators in the three-tank model for resulting the FDI residuals and for detecting the faults working around some various desired points in turn. In addition, modern fuzzy and NN control on the tanks have been studied by \cite{R11,R12,R13}. However, this deterministic scenarios ignores the uncertainties and disturbances of the system which could be ameliorated by its counterpart stochastic approach.

There have been several the well-known recursive classic Kalman techniques dealing with the non-linear problem, comprising from the extended, the unscented to the Gaussian Kalman filter \cite{R14}. Among those algorithms, the extended Kalman filtering (EKF) is a stochastic method being used to deal with the non-linear problem \cite{R15}. Due to the unknown stochastic-noise correlation in EKF leading to lack of mathematical corresponding between the Kalman-type output and the noise covariance, the critical proper choices of noises in both system-$\mathbf{Q}$ and measurement-$\mathbf{R}$ to obtain the desired performance working inside the error threshold are mathematically not fulfilled. To cope with, the two static and dynamic matrix mechanism have been proposed. This innovative matrix of system noise $\mathbf{Q}$ guarantee the full couple of variance and covariance but no literature has implemented this into the three-tank process. This is due to the fact that this dynamic-noise Kalman (AKF) is tremendously 
applied in the navigation (INS) and the positioning (GPS) by \cite{R16,R17,R18,Wafi-AIP}.

Regarding the problem formulation, This paper is started with the mathematical description. Moreover, the linear control design and the non-linear with stabilization and exact linearization are studied in the following to examine the performance of reference tracking. Lastly, the innovative noise (AKF) algorithm is implemented to estimate the true system ended by some conlusion and future work related to the paper.

\section{Mathematical Descriptions}\label{C2}
\begin{figure*}[h!]
    \centering
    \includegraphics[width=0.75\linewidth]{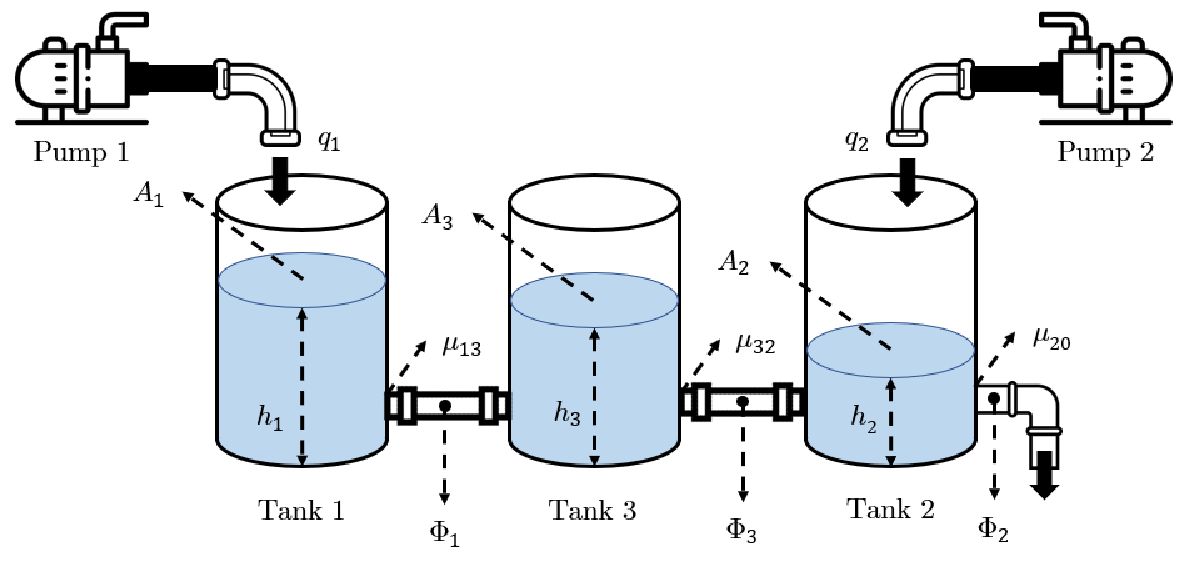}
    \caption{Three-tank hydraulic dynamical-system}
    \label{F1}
\end{figure*}

The proposed three-tank hydraulic system is illustrated in Fig.~(\ref{F1}). 
The system consists of three identical cylindrical tanks arranged in parallel, 
each having the same cross-sectional area $A_n$, where $\tau = 3$ and 
$n = 1, \dots, \tau$, such that $A_1 = A_2 = A_3$.
The tanks are interconnected through three pipes: between tanks 1 and 3 $(\Phi_1)$, 
between tanks 3 and 2 $(\Phi_3)$, and an outlet pipe $(\Phi_2)$. 
All pipes share the same cross-sectional area, i.e., 
$\Phi_1 = \Phi_2 = \Phi_3 = \Phi$. 
However, the outflow coefficients $(\mu_{ab})$ associated with these pipes are not identical, 
where $\mu_{13} = \mu_{32} \neq \mu_{20}$.

Tanks 1 and 2 are actuated by two DC motor-driven pumps that provide inlet flow rates 
$q_1$ and $q_2$, respectively. These flow rates are expressed in $\text{m}^3\text{s}^{-1}$ 
and are controlled via a digital-to-analog (D/A) converter. 
Each pump has a maximum flow capacity denoted by $q_m^\ast$, for $m \in \{1,2\}$.
The liquid levels in the tanks, denoted by $h_n$, are measured using differential pressure 
piezo-resistive transducers, which convert the fluid levels into voltage signals. 
The maximum allowable level in each tank is denoted by $h_n^\ast$.
\begin{figure*}[h!]
    \centering
    \includegraphics[width=0.85\linewidth]{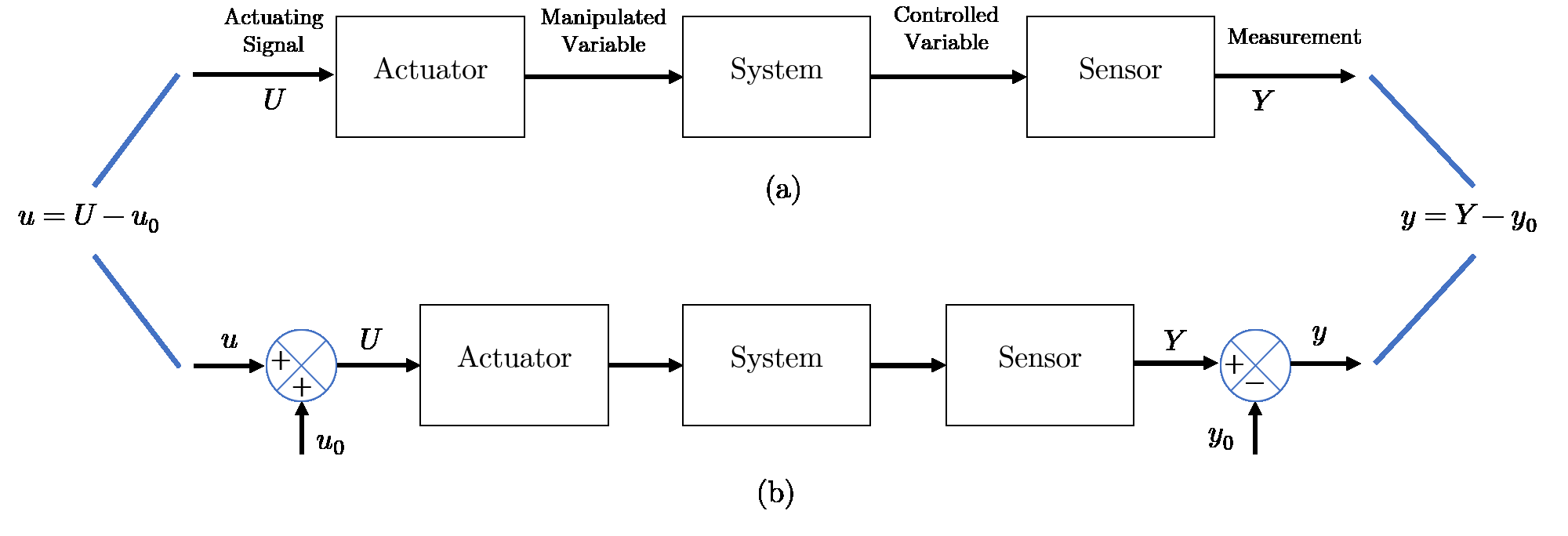}
    \includegraphics[width=0.65\linewidth]{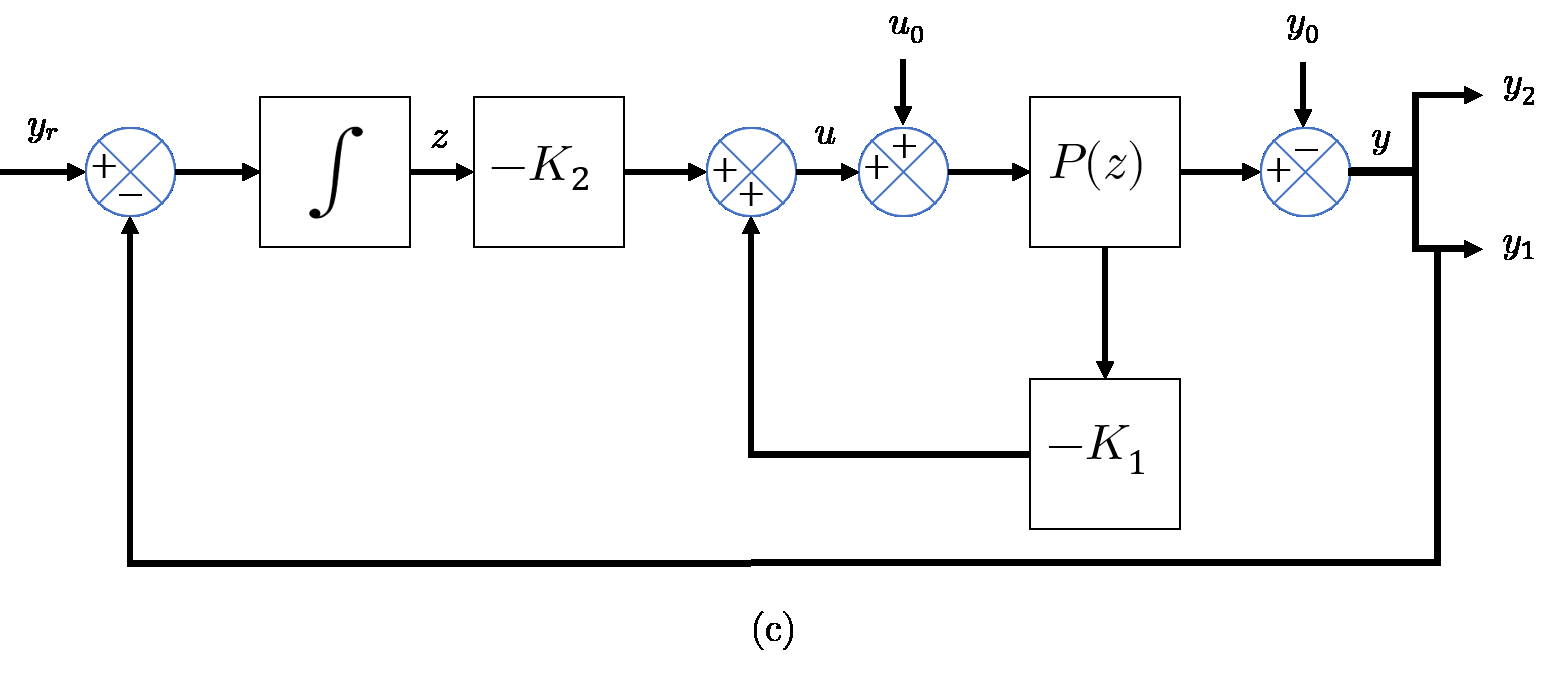}
    \caption{(a) The dynamic model representation taking into account the global operating points; (b) The linearized model representation considering the operating point $(u_0,y_0)$; (c) The basic nominal tracking control concept working around the operating point $(u_0,y_0)$}
    \label{F2}
\end{figure*}

Furthermore, the rate of change of the total mass $M_t$ in the system is determined by 
the difference between the inlet mass flow rate $M_i$ and the outlet mass flow rate $M_o$, 
as described by Bernoulli's principle:
\begin{align}
    \frac{dM_t}{dt} = M_i - M_o,
    \label{Eq1}
\end{align}
which can be reformulated in terms of fluid density $\rho_n$ as:
\begin{align}
    A \dot{h} = \rho q_i - \rho q_o.
    \label{Eq2}
\end{align}

Assuming the fluid is water, the density is constant across all tanks, 
i.e., $\rho_1 = \rho_2 = \rho_3$. 
Thus, the dynamic behavior of each tank depends on the net inflow and outflow, given by:
\begin{align}
    A_n \frac{dh_n}{dt} = \sum_{j \in \mathcal{N}} (q_i)_n - \sum_{k \in \mathcal{N}} (q_o)_n.
    \label{Eq3}
\end{align}
For inter-tank flow (excluding pump inputs), the flow rate from tank $a$ to tank $b$ is 
denoted by $q_{ab}$, where $a,b = 1, \dots, \tau$ and $a \neq b$. 
According to Torricelli’s law, this flow is expressed as:
\begin{align}
    q_{ab}(t) = \mu_{ab} \Phi \, \text{sign}(h_a(t) - h_b(t)) 
    \sqrt{2g \, |h_a(t) - h_b(t)|}.
    \label{Eq4}
\end{align}
The only external outflow from the system occurs from tank 2, given by:
\begin{align}
    q_{20}(t) = \mu_{20} \Phi \sqrt{2g h_2(t)}.
    \label{Eq5}
\end{align}

Based on the above relations, the nonlinear mass balance equations for the three-tank system are:
\begin{subequations}
\begin{align}
    A \frac{dh_1(t)}{dt} &= q_1(t) - q_{13}(t), \label{Eq6a}\\
    A \frac{dh_2(t)}{dt} &= q_2(t) + q_{32}(t) - q_{20}(t), \label{Eq6b}\\
    A \frac{dh_3(t)}{dt} &= q_{13}(t) - q_{32}(t). \label{Eq6c}
\end{align}
\end{subequations}

\subsection{Linear Representation}\label{C2a}

A linear representation of the three-tank system is obtained by linearizing the 
nonlinear model around an equilibrium point $(u_0,y_0)$ under the operating 
condition $h_1 > h_3 > h_2$. This assumption determines the flow directions among 
the tanks and allows the nonlinear hydraulic relations to be locally approximated 
by a first-order Taylor expansion.

Next, let the state vector, input vector, and output vector be defined as $x = [h_1,h_2,h_3]^\top\in\mathbb{R}^3$, $u = [q_1,q_2]^\top\in\mathbb{R}^2$, and $y = [h_1,h_2,h_3]^\top\in\mathbb{R}^3$.
The equilibrium point is denoted by $(u_0,y_0)$, where $u_0$ is the constant input 
that maintains the system at the steady-state level $y_0$. Around this operating point, 
the variables are written in deviation form as
\begin{align}
    u = U - u_0, \qquad y = Y - y_0,
    \label{Eq9}
\end{align}
where $U$ and $Y$ denote the actual plant input and output, respectively. For clarity, consider the system in Fig.~(\ref{F2}a) with input $U$ and output $Y$, and its linearized 
representation around the equilibrium point shown in Fig.~(\ref{F2}b).
Thus, the linearized model describes only small deviations from the equilibrium point, 
rather than the absolute variables themselves.
\begin{figure*}[b!]
\hrule
\begin{align}
    F = \begin{bmatrix}
    -\dfrac{\mu_{13}\Phi\varphi}{2A\sqrt{y_{01}-y_{03}}} & 0 & \dfrac{\mu_{13}\Phi\varphi}{2A\sqrt{y_{01}-y_{03}}}\\[1em]
    0 & -\dfrac{\mu_{13}\Phi\varphi}{2A\sqrt{y_{01}-y_{03}}} - \dfrac{\mu_{20}\Phi\varphi}{2A\sqrt{y_{02}}} & \dfrac{\mu_{32}\Phi\varphi}{2A\sqrt{y_{03}-y_{02}}}\\[1em]
    -\dfrac{\mu_{13}\Phi\varphi}{2A\sqrt{y_{01}-y_{03}}} & \dfrac{\mu_{32}\Phi\varphi}{2A\sqrt{y_{03}-y_{02}}} & -\dfrac{\mu_{32}\Phi\varphi}{2A\sqrt{y_{03}-y_{02}}} + \dfrac{\mu_{13}\Phi\varphi}{2A\sqrt{y_{01}-y_{03}}}
    \end{bmatrix},
    \quad
    B = \begin{bmatrix}
    \dfrac{1}{A} & 0\\[0.75em]
    0 & \dfrac{1}{A}\\[0.75em]
    0 & 0
    \end{bmatrix},
    \label{Eq8}
\end{align}
\end{figure*}

By applying first-order Taylor expansion to the nonlinear model, the continuous-time 
linearized system is written as
\begin{align}
    \dot{x}(t) = F x(t) + B u(t),
\end{align}
where $F$ is the state matrix obtained from the Jacobian of the nonlinear dynamics 
with respect to the state variables, and $B$ is the input matrix derived from the 
Jacobian with respect to the inputs. The matrix $F$ captures the interactions among 
tank levels around the operating point, while $B$ describes the influence of the pump 
flow rates on the system dynamics. The matrices $F$ and $B$ are given in (\ref{Eq8}), 
with $\varphi = \sqrt{2g}$.

Since the control law and the Kalman estimator are implemented digitally, the model 
must be expressed in discrete time. For this reason, the continuous-time linearized 
system is discretized using a sampling period $t_s$. The resulting linear time-invariant 
(LTI) discrete-time model is written as
\begin{align}
\left\{
\begin{aligned}
    x(k+1) &= A_d x(k) + B_d u(k), \\
    y(k) &= C x(k),
\end{aligned}
\right.
\label{Eq7}
\end{align}
where $k$ is the discrete sampling instant.
Here, $A_d \in \mathbb{R}^{n \times n}$ is the discrete-time state-transition matrix, 
and $B_d \in \mathbb{R}^{n \times m}$ is the discrete-time input matrix. The matrix 
$C \in \mathbb{R}^{q \times n}$ denotes the output matrix selecting the measured states. 
Since all tank levels are measurable, $C$ is chosen as $C = I_3$. 

The discrete-time matrices $A_d$ and $B_d$ are obtained from the continuous-time 
matrices $F$ and $B$ via discretization with sampling time $t_s$. Under a zero-order 
hold assumption,
\begin{align}
    A_d &= e^{F t_s}, \qquad 
    B_d = \int_0^{t_s} e^{F\tau} B \, d\tau.
\end{align}
This discrete-time model is required for the sampled-data implementation of both 
the controller and the Kalman filter.

The choice of the operating point remains important, since the linearized model is only 
valid locally around $(u_0,y_0)$. Hence, the matrices $F$, $B$, $A_d$, and $B_d$ 
represent the system dynamics only in the neighborhood of the selected equilibrium.

\subsection{Nominal Tracking Control Law}\label{C2b}

According to \cite{R20}, the number of controlled variables $\psi_y$ must not exceed 
the number of control inputs $\psi_u$, i.e., $\psi_y \leq \psi_u$. Otherwise, a subset 
of the available states must be selected to track the reference signal $y_r$, as 
illustrated in Fig.~(\ref{F2}c). 

Since the system has two inputs $(q_1, q_2)$, two output variables are selected for 
tracking, defined as
\begin{align}
    y(k) = \begin{bmatrix}
    y_1(k)\\
    y_2(k)
    \end{bmatrix} = \begin{bmatrix}
    C_1 \\ C_2
    \end{bmatrix} x(k),
    \qquad
    y_1 = [ h_1, h_2 ]^\top,
    \label{Eq10}
\end{align}
where $y_2 = h_3$, and $y_1 \in \mathbb{R}^p$ with $p \leq m$. The control objective 
is to achieve zero steady-state error, i.e., $y_r - y_1 = 0$.

\begin{figure}[h!]
    \centering
    \includegraphics[width=\linewidth]{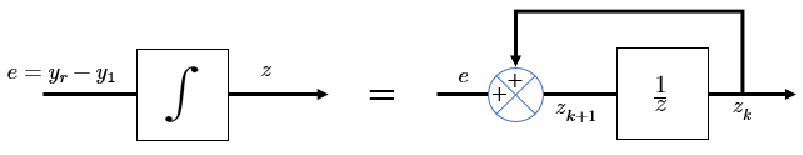}
    \caption{Integrator definition on Fig.~(\ref{F2}c)}
    \label{F3}
\end{figure}

To this end, an integral action is introduced through a comparator–integrator pair, 
given by
\begin{align}
    z_m(k+1) = z_m(k) + t_s \big( y_{r,m}(k) - y_{1,m}(k) \big),
    \label{Eq11}
\end{align}
where $y_1 = C_1 x$ and $z \in \mathbb{R}^p$ with 
$z = [z_1, z_2]^\top$. 
This integrator, illustrated in Fig.~(\ref{F3}), accumulates the tracking error to 
eliminate steady-state offsets. The sampling time $t_s$ must be properly selected to 
ensure closed-loop stability. 

Next, the augmented state vector is therefore defined as $x_e = [ x^\top, z^\top ]^\top$, 
leading to the extended state-space model:
\begin{align}\left\{
    \begin{aligned}
        \begin{bmatrix}x(k+1)\\
        z(k+1)\end{bmatrix} =& \begin{bmatrix}A_d & \textbf{O}_{n,p}\\
        -t_sC_1 & I_p\end{bmatrix}\begin{bmatrix}x(k)\\
        z(k)\end{bmatrix} + \begin{bmatrix}B_d\\
        \textbf{O}_{p,m}\end{bmatrix}u \\
        & + \begin{bmatrix}\textbf{O}_{n,p}\\
        t_sI_p\end{bmatrix}y_r\\
        y(k) =& \begin{bmatrix}C & \textbf{O}_{q,p}\\
        \end{bmatrix}\begin{bmatrix}x(k)\\
        z(k)\end{bmatrix}
    \end{aligned}\right. \label{Eq12}
\end{align}
where $I_v$ denotes the $v$-dimensional identity matrix and $\mathbf{O}_{n,p}$ is a 
zero matrix of size $n \times p$.

For simplicity, (\ref{Eq12}) can be compactly written as
\begin{align}
\left\{
\begin{aligned}
    x_e(k+1) &= \bar{A}_d x_e(k) + \bar{B}_d u(k) + B_r y_r(k), \\
    y(k) &= \bar{C} x_e(k),
\end{aligned}
\right.
\label{Eq13}
\end{align}
where $\bar{A}_d$, $\bar{B}_d$, and $\bar{C}$ are the augmented system matrices.

The control law is then designed using pole placement, with desired eigenvalues 
$\lambda$, as
\begin{align}
    u(k) = -K x_e(k) = -\begin{bmatrix} K_1 & K_2 \end{bmatrix}
    \begin{bmatrix}
    x(k)\\
    z(k)
    \end{bmatrix}.
    \label{Eq14}
\end{align}

\subsection{Non-linear Decoupling Control Law}\label{C2c}

Although the linear state-feedback controller achieves satisfactory tracking, 
it is valid only near the operating point due to the linearization assumption. 
Since the three-tank system is inherently nonlinear, performance may degrade 
under large operating variations. 
To address this limitation, a nonlinear control approach based on exact feedback 
linearization and decoupling is employed, enabling independent control of each 
output while preserving the nonlinear dynamics.

The system in (\ref{Eq7}) can be written in its nonlinear form as
\begin{align}
\left\{
\begin{aligned}
    \dot{x}(t) &= \Delta(x(t)) + \sum_{i=1}^m \xi_i(x(t)) u_i(t), \quad m \in [1,2],\\
    y(t) &= H(x(t)),
\end{aligned}
\right.
\label{Eq15}
\end{align}
where $\Delta(x)$ represents the nonlinear drift dynamics derived from the mass 
balance equations, while $\xi_i(x)$ denote the input vector fields. Specifically,
\begin{align}
    \Delta(\cdot) = \frac{1}{A}
    \begin{bmatrix}
    -q_{13}(t)\\
    q_{32}(t) - q_{20}(t)\\
    q_{13}(t) - q_{32}(t)
    \end{bmatrix}, 
    \quad
    \xi =
    \begin{bmatrix}
    \frac{1}{A} & 0 & 0\\
    0 & \frac{1}{A} & 0
    \end{bmatrix}^\top.
    \label{Eq16}
\end{align}
The control objective is to track a reference signal $y_r$, similar to the linear case. 
Therefore, an exact feedback linearization with decoupling is employed using the control law
$u = \alpha(x) + \beta(x)\zeta,$
assuming equal input-output dimensions $(q = m)$.
The relative degree $\vartheta_i$ for each output is defined as
\begin{align}
    \vartheta_i = \left\{ \min l \in \mathbb{N} \; \middle| \; \exists j \in [1:m],\ 
    L_{\xi_j} L_\Delta^{l-1} H_i(x) \neq 0 \right\}.
    \label{Eq17}
\end{align}
Here, $L$ denotes the Lie derivative. In particular,
\begin{align}
    L_\Delta H_i(x) = \sum_{j=1}^n \Delta_j(x)\frac{\partial H_i}{\partial x_j}(x),
    \label{Eq18}
\end{align}
with the recursive definition
\begin{align}
\left\{
\begin{aligned}
    L_\Delta^0 H(x) &= H(x),\\
    L_\Delta^k H(x) &= L_\Delta \big( L_\Delta^{k-1} H(x) \big), \quad k \geq 1.
\end{aligned}
\right.
\label{Eq19}
\end{align}

If the relative degrees $\vartheta_i$ exist for all $i \in [1:m]$, the decoupling matrix is defined as
\begin{align}
    \Lambda(x) =
    \begin{bmatrix}
    L_{\xi_1} L_\Delta^{\vartheta_1 - 1} H_1(x) & \dots & L_{\xi_m} L_\Delta^{\vartheta_1 - 1} H_1(x)\\
    \vdots & \ddots & \vdots\\
    L_{\xi_1} L_\Delta^{\vartheta_m - 1} H_m(x) & \dots & L_{\xi_m} L_\Delta^{\vartheta_m - 1} H_m(x)
    \end{bmatrix},
    \label{Eq20}
\end{align}
with
\begin{align}
    \Lambda_0(x) =
    \begin{bmatrix}
    L_\Delta^{\vartheta_1} H_1(x)\\
    \vdots\\
    L_\Delta^{\vartheta_m} H_m(x)
    \end{bmatrix}.
    \label{Eq21}
\end{align}

The system is input-output decouplable on a subset $\Pi \subset \mathbb{R}^n$ if 
$\mathrm{rank}(\Lambda(x)) = m$ for all $x \in \Pi$ \cite{R21}. 
The nonlinear control law is then given by
\begin{align}
    u(t) = \underbrace{-\Lambda^{-1}(x)\Lambda_0(x)}_{\alpha(x)} + \underbrace{\Lambda^{-1}(x)}_{\beta(x)}\zeta(t) \label{Eq22}
\end{align}
which yields the linearized input-output relation
\begin{align}
    y_i^{(\vartheta_i)}(t) = \zeta_i(t), \quad \forall i \in [1:m].
    \label{Eq23}
\end{align}

For the present system, $\sum_{i=1}^m \vartheta_i = 2 < n = 3$, indicating the presence 
of internal (unobservable) dynamics associated with tank 3. However, this internal 
dynamics is stable, allowing the linearized control law to be applied.
By contrast, if $\sum_{i=1}^m \vartheta_i = n$, the system is fully 
input--output linearizable, implying the absence of internal dynamics 
and yielding a completely decoupled linear system.

The resulting decoupled subsystems behave as integrators:
\begin{align}
    \frac{h_1(s)}{\zeta_1(s)} = \frac{h_2(s)}{\zeta_2(s)} = \frac{1}{s},
    \label{Eq24}
\end{align}
where $(s)$ is the Laplace variable. Since these subsystems are marginally stable, 
an additional proportional feedback is introduced:
\begin{align}
    \zeta_i(t) = K_i \big( y_{r,i}(t) - h_i(t) \big), \quad \forall i \in [1:m],
    \label{Eq25}
\end{align}
leading to the closed-loop transfer function
\begin{align}
    \frac{h_i(s)}{y_{r,i}(s)} = \frac{K_i}{s + K_i}.
    \label{Eq26}
\end{align}

\begin{figure*}[t!]
    \centering
    \includegraphics[width=.95\linewidth]{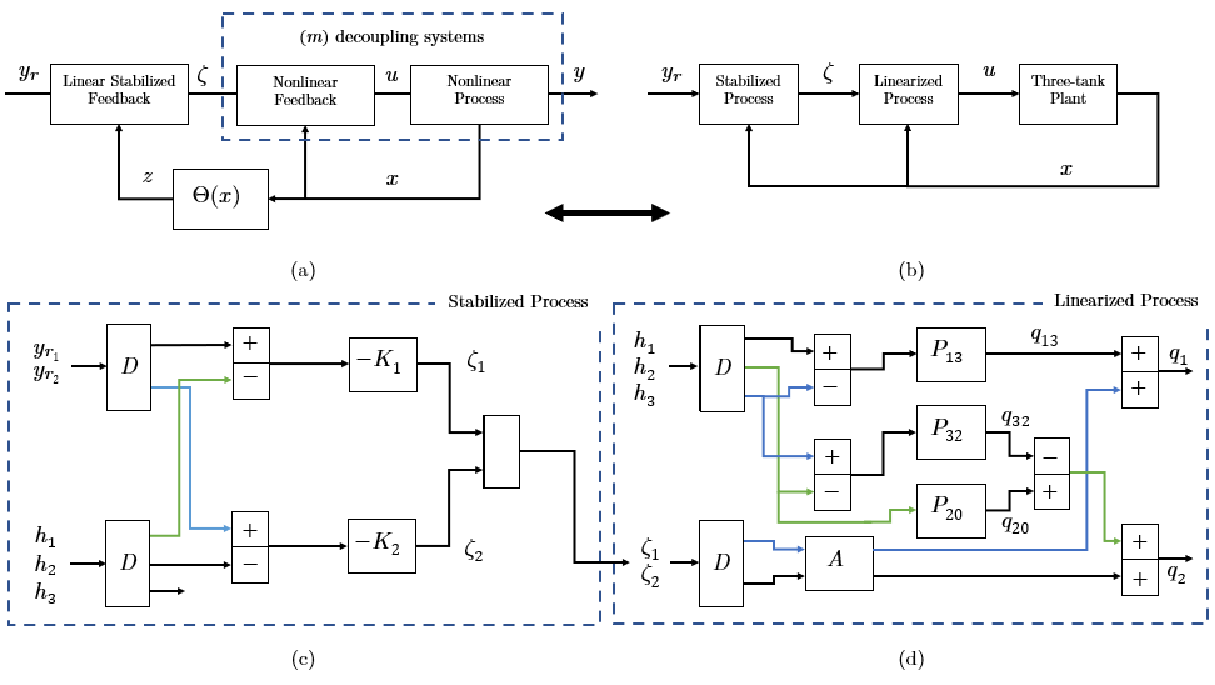}
    \caption{(a) The basic control law of non-linear system; (b) The generalized non-linear control fault-free scenario; (c) The stabilized block diagram for non-linear dynamical system; (d) The linearized block diagram for non-linear dynamical system}
    \label{F4}
\end{figure*}

The overall control structure is illustrated in Fig.~(\ref{F4}), where Fig.~(\ref{F4}c) 
shows the stabilized configuration and Fig.~(\ref{F4}d) the linearized control scheme.

\section{Adaptive Covariance Kalman Filter}\label{C3}

The proposed adaptive covariance scheme is motivated by recent advances in 
estimation and control of nonlinear and interconnected systems. In particular, 
distributed estimation \cite{Add1,Add2,Wafi-QuadrupleTank} and decentralized control \cite{Add3,Add4,Wafi-MRAC} strategies for multi-tank processes, 
as well as fault-tolerant control \cite{Add5,Wafi-Scrubber,Add6} under sensor uncertainties, highlight the need 
for reliable and adaptive state estimation. Moreover, robust and non-fragile 
estimation-based control methods have shown that fixed noise covariance assumptions 
may degrade performance in the presence of uncertainties \cite{Wafi-Elham}. 

Therefore, this section proposes an adaptive covariance Kalman filter (AKF) based 
on the conventional centralized Kalman filter (CKF) applied to the system dynamics 
in (\ref{Eq8}) \cite{R23}, 
which improves estimation accuracy by dynamically updating 
the process noise covariance without requiring precise prior knowledge of system noise.

The linearized state-transition matrix is given by
\begin{align}
    \mathbf{F}_{k|k-1} = \left.\frac{\partial \mathbf{f}_e(\mathbf{x},\mathbf{u}_k)}{\partial \mathbf{x}}\right|_{\mathbf{x} = \mathbf{x}_{k-1}},
    \label{Eq27}
\end{align}
where $\mathbf{f}_e(\cdot) = A_d\mathbf{x} + B_d\mathbf{u}_k$.
The CKF recursion is defined as
\begin{align}\label{Eq28}
    \hat{\mathbf{x}}_k^{+} &= \mathbf{f}_e\left(\hat{\mathbf{x}}_{k-1},\mathbf{u}_k\right),\\
    \mathbf{P}_k^{+} &= \mathbf{F}_{k|k-1}\mathbf{P}_{k-1}\mathbf{F}_{k|k-1}^\top + \mathbf{Q}_{k-1},\\
    \mathbf{K}_k &= \mathbf{P}_{k}^{+}\mathbf{H}^\top \left(\mathbf{H}\mathbf{P}_k^{+}\mathbf{H}^\top + \mathbf{R}_k\right)^{-1},\\
    \hat{\mathbf{x}}_k &= \hat{\mathbf{x}}_k^{+} + \mathbf{K}_k\left(\mathbf{y}_k - \mathbf{H}\hat{\mathbf{x}}_k^{+}\right),\\
    \mathbf{P}_k &= \left(\mathbf{I} - \mathbf{K}_k\mathbf{H}\right)\mathbf{P}_k^{+}.
\end{align}

Here, $\mathbf{P}_k^{+}$ and $\mathbf{P}_k$ denote the predicted and updated error 
covariance matrices, respectively, $\mathbf{K}_k$ is the Kalman gain, $\mathbf{Q}_k$ 
and $\mathbf{R}_k$ are the process and measurement noise covariance matrices, and 
$\mathbf{I}$ is the identity matrix.
To enhance estimation performance, the process noise covariance $\mathbf{Q}_k$ is 
adaptively updated while assuming $\mathbf{R}_k$ is known \cite{R22}. The innovation 
covariance is estimated as
\begin{align}
    \hat{\mathbf{C}}_{\mathbf{v}_k} = \frac{1}{\Psi}\sum\nolimits_{i=i_0}^k \mathbf{v}_i\mathbf{v}_i^\top,
    \qquad
    \mathbf{v}_k = \mathbf{y}_k - \mathbf{H}\hat{\mathbf{x}}_k,
    \label{Eq29}
\end{align}
where $\mathbf{v}_k$ is the innovation sequence and $i_0 = k - \Psi + 1$ defines 
the window length $\Psi$.

The adaptive update of $\mathbf{Q}_k$ is given by
\begin{align}
    \hat{\mathbf{Q}}_k = Q_\alpha + \mathbf{P}_k - \mathbf{F}_{k|k-1}\mathbf{P}_{k-1}\mathbf{F}_{k|k-1}^\top,
    \label{Eq30}
\end{align}
where $\Delta\mathbf{x}_k = \hat{\mathbf{x}}_k - \hat{\mathbf{x}}_k^{+}$. 
Under a steady-state approximation, this becomes
\begin{align}
    \hat{\mathbf{Q}}_k \approx 
    \underbrace{\frac{1}{\Psi}\sum\nolimits_{i=i_0}^k \Delta\mathbf{x}_i\Delta\mathbf{x}_i^\top}_{Q_\alpha}
    \approx \mathbf{K}_k\hat{\mathbf{C}}_{\mathbf{v}_k}\mathbf{K}_k^\top.
\end{align}

\begin{figure}[t!]
    \centering
    \includegraphics[width=.8075\linewidth]{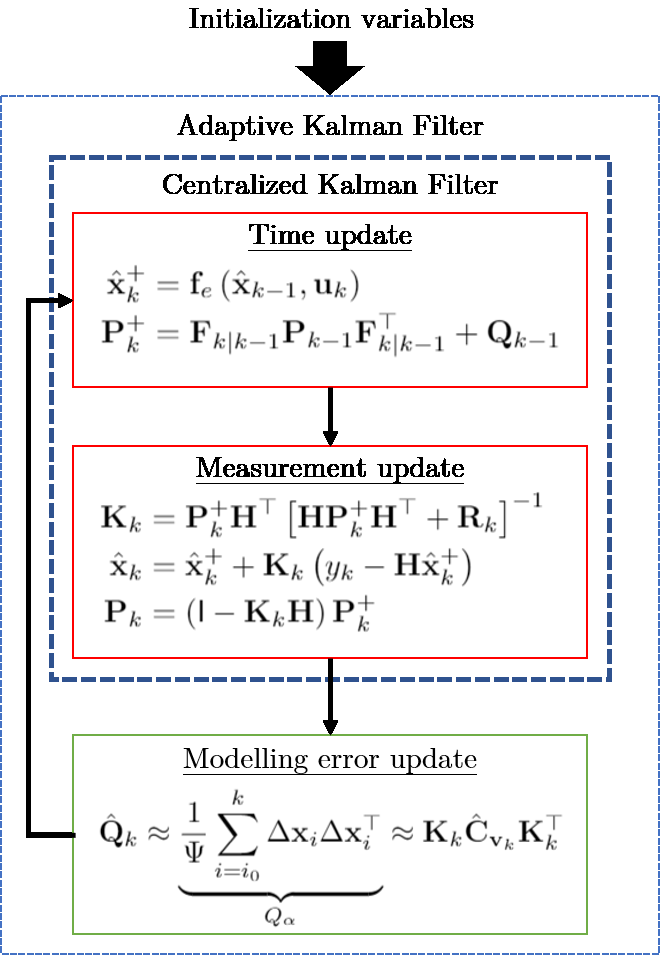}
    \caption{Adaptive $\mathbf{Q}$ update mechanism (AKF) based on CKF}
    \label{F5}
\end{figure}

The stability properties of the estimator are discussed in \cite{Wafi-Hydraulic,Wafi-SysID}. 
They rely on the boundedness of $\mathbf{Q}$ and $\mathbf{R}$ within the intervals 
$(q^{-},q^{+})$ and $(r^{-},r^{+})$, respectively. In addition, the system must be 
observable, i.e., $(A_d,\mathbf{H})$ is observable, and the initial covariance must 
satisfy $\mathbf{P}_0 > 0$. Under these conditions, the Riccati equation remains bounded:
\begin{align}
    q^{-}\mathbf{I} \leq \mathbf{P}_k \leq q^{+}\mathbf{I}, \quad \forall k \geq 0.
\end{align}

\section{Numerical Simulation Results}\label{C4}

This section presents three simulation scenarios to evaluate the proposed methods. 
The system parameters are listed in Table~(\ref{T1}), with sampling time $t_s = 1\,\text{s}$, 
under the assumptions described in Section~\ref{C2b}.
\begin{table}[h!]
    \centering
    \caption{Three-tank process parameters}
    \begin{tabular}{c|c|c}
        \toprule
        Variable & Symbol & Value \\
        \midrule
        CSA$^\ast$ of tank & A & 0.0154 $m^2$\\
        CSA$^\ast$ of inter-tank & $\Phi$ & $5\times 10^{-5}$ $m^2$ \\
        Coefficient of outflow & $\mu_{13} = \mu_{32}$ & 0.5 \\
        Coefficient of outflow & $\mu_{20}$ & 0.675 \\
        Peak of flow-rate & $q_m^\ast (m\in[1,2])$ & $1.2\times 10^{-4}$ $m^3s^{-1}$ \\
        Peak of level & $h_n^\ast (n\in[1:\tau])$ & 0.62 $m$ \\
        \bottomrule
        \multicolumn{3}{r}{$^\ast$Note: CSA means cross-sectional area} 
    \end{tabular}
    \label{T1}
\end{table}

The control objective is to regulate the plant around the operating point $(u_0,y_0)$, given by
\begin{align*}
    u_0 &= \begin{bmatrix}
    0.35\times 10^{-4} & 0.375\times 10^{-4}
    \end{bmatrix}^\top \; (m^3/s), \\
    y_0 &= \begin{bmatrix}
    0.40 & 0.20 & 0.30
    \end{bmatrix}^\top \; (m).
\end{align*}

The feedback gain is designed using pole placement with desired eigenvalues
\begin{align*}
    \lambda &= \begin{bmatrix}
    0.92 & 0.97 & 0.90 & 0.95 & 0.94
    \end{bmatrix},
\end{align*}
resulting in
\begin{align*}
    \begin{bmatrix} K_1 & K_2 \end{bmatrix}
    = 10^{-4}
    \begin{bmatrix}
    21.6 & 3 & -5 & -0.95 & -0.32\\
    2.9 & 19 & -4 & -0.30 & -0.91
    \end{bmatrix}.
\end{align*}
\begin{figure*}[t!]
	\centering
	\begin{subfigure}[t]{0.425\linewidth}
		\includegraphics[width=\linewidth]{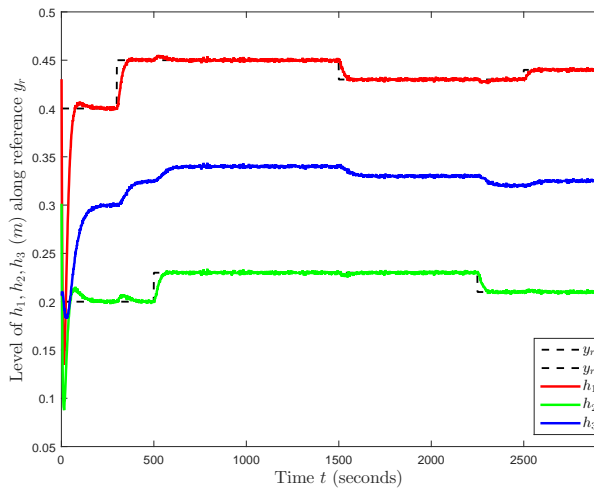}
		\caption{}
		\label{F6a}
	\end{subfigure}\qquad
    \begin{subfigure}[t]{0.425\linewidth}
		\includegraphics[width=\linewidth]{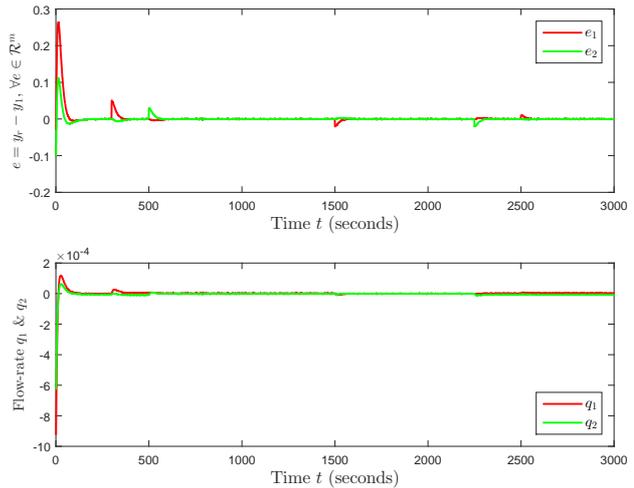}
		\caption{}
		\label{F6b}
	\end{subfigure}
    \caption{(a) The tracking performance of the linear control law over some dynamic references; (b) The error $(e_i)$ and the control input $(q_i)$ of the respected level}
    \label{F6}
\end{figure*}

The results are presented in Fig.~(\ref{F6}), illustrating the effectiveness of the 
linear control design. The system outputs $y$ (colored lines) are compared with the 
dynamic step references $y_r$ (black dashed lines). This MIMO system has two 
input-output pairs $(m = 2)$, where $h_1 \rightarrow y_{r,1}$ and $h_2 \rightarrow y_{r,2}$, 
operating around the equilibrium point $(u_0,y_0)$.
As shown in Fig.~(\ref{F6a}), the control law successfully tracks both reference 
signals $y_{r,i}$. However, due to the coupling between the tanks, changes in one 
level $(h_i)$ influence the dynamics of the other level $(h_j)$, which is reflected 
by the slight interaction observed in the transient responses. 
Fig.~(\ref{F6b}) shows the corresponding tracking errors $(e_i)$ and the control 
inputs $(q_i)$ generated by the pumps, indicating that the control action remains 
within acceptable bounds while achieving the desired tracking performance.

For the nonlinear case (Section~\ref{C2c}), the system has two outputs with relative 
degrees $\vartheta_1$ and $\vartheta_2$. Since $\mathrm{rank}(\Lambda(x)) = m$, the system 
is fully decouplable. The matrices $\Lambda(x)$ and $\Lambda_0(x)$ are obtained as
\begin{gather*}
    \Lambda(x) =
    \begin{bmatrix}
    \frac{1}{A} & 0\\
    0 & \frac{1}{A}
    \end{bmatrix}, \quad
    \Lambda_0(x) =
    \begin{bmatrix}
    -\frac{1}{A}q_{13}\\
    \frac{1}{A}(q_{32} - q_{20})
    \end{bmatrix},
\end{gather*}
leading to the control input
\begin{align*}
    u(t) =
    \frac{1}{A}
    \begin{bmatrix}
    -q_{13}(x_t)\\
    q_{32}(x_t) - q_{20}(x_t)
    \end{bmatrix}
    +
    \begin{bmatrix}
    A & 0\\
    0 & A
    \end{bmatrix}
    \zeta(t).
\end{align*}

\begin{figure*}[t!]
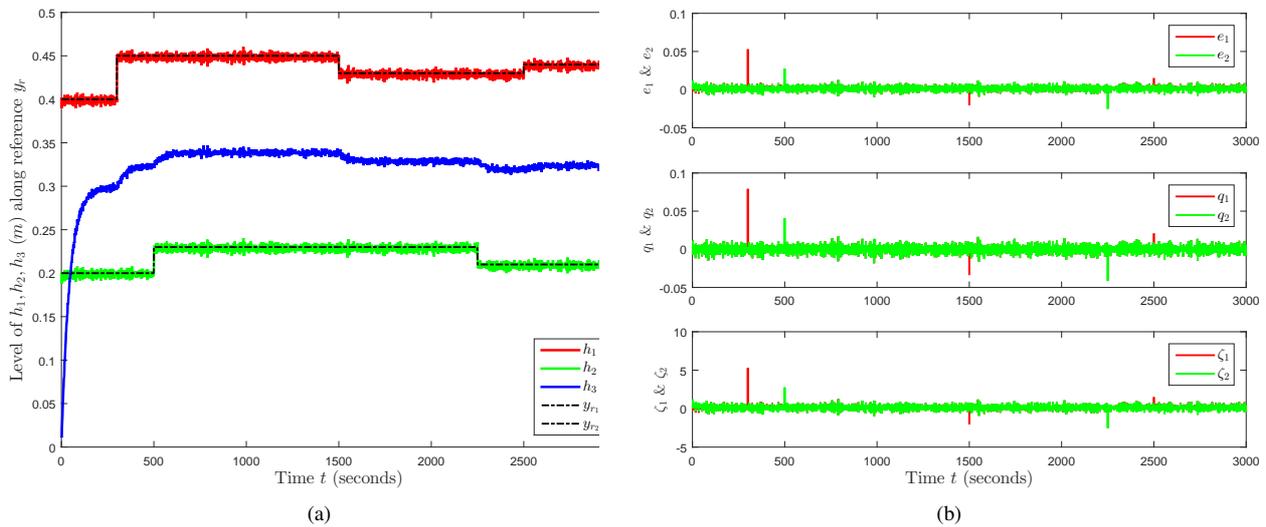

	\centering
	\begin{subfigure}[t]{0.425\linewidth}
		\includegraphics[width=\linewidth]{Graph/Fig3.eps}
		\caption{}
		\label{F7a}
	\end{subfigure}\qquad
    \begin{subfigure}[t]{0.425\linewidth}
		\includegraphics[width=\linewidth]{Graph/Fig4.eps}
		\caption{}
		\label{F7b}
	\end{subfigure}
    \caption{(a) The tracking performance of the non-linear control law with the stabilized and linearized decoupling control law over some dynamic references; (b) The error $(e_i)$, the control input $(q_i)$ and the equal linearized input $(\zeta_i)$ of the respected level}
    \label{F7}
\end{figure*}

The nonlinear responses are shown in Fig.~(\ref{F7}). The outputs track the references 
with improved decoupling behavior, indicating that variations in one tank do not 
significantly affect the others. Fig.~(\ref{F7b}) shows the tracking errors, control 
inputs $(q_i)$, and auxiliary inputs $(\zeta_i)$. Compared to the linear case, the 
nonlinear control achieves better decoupling performance, although measurement noise 
has a more noticeable effect on the outputs.

For the AKF, the initial conditions are set as
\begin{gather*}
    \mathbf{x}_0 =
    \begin{bmatrix}
    0.9 & 0.55 & 0.5
    \end{bmatrix}^\top, \\
    \mathbf{P}_0 = 10 I_3, \quad
    \mathbf{Q}_0 = 10^{-12} I_3, \quad
    \mathbf{R}_0 = 5^2.
\end{gather*}

\begin{figure*}[t!]
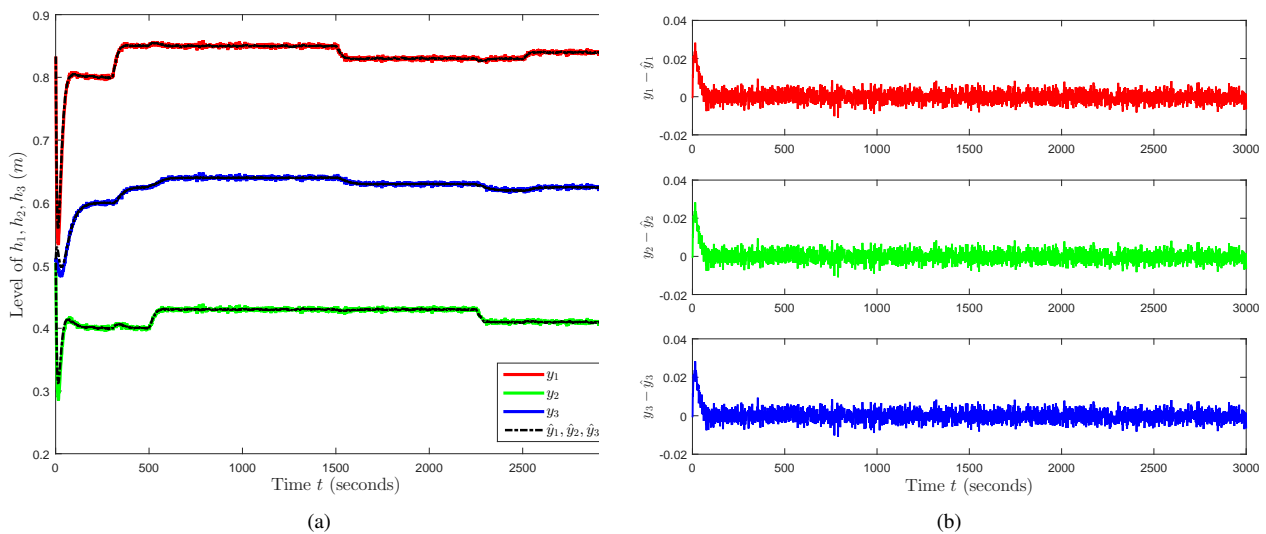

	\centering
	\begin{subfigure}[t]{0.425\linewidth}
		\includegraphics[width=\linewidth]{Graph/Fig5.eps}
		\caption{}
		\label{F8a}
	\end{subfigure}\qquad
    \begin{subfigure}[t]{0.43\linewidth}
		\includegraphics[width=\linewidth]{Graph/Fig6.eps}
		\caption{}
		\label{F8b}
	\end{subfigure}
    \caption{(a) The estimation performance $(\hat{y}_i)$ of AKF over the output systems $(y_i)$; (b) The estimation error $(\hat{e}_i)$ from the true values of the respected level}
    \label{F8}
\end{figure*}

The estimation performance is shown in Fig.~(\ref{F8}). The estimated outputs $(\hat{y}_i)$ 
closely follow the true states $(y_i)$, while the estimation errors remain small and bounded. 
These results demonstrate the effectiveness of the adaptive covariance mechanism in improving 
state estimation accuracy under nonlinear dynamics.


\section{Conclusion}

This paper presented the mathematical modeling of a three-tank system together 
with linear state-feedback control and nonlinear decoupling control based on 
exact feedback linearization. Both control strategies demonstrated effective 
tracking of the reference signals $(y_r)$ for tanks 1 and 2. In addition, the 
proposed adaptive covariance Kalman filter (AKF) showed improved estimation 
performance under nonlinear dynamics, providing accurate and reliable state 
estimates.
For future work, the framework can be extended to more complex systems with 
unobservable or non-full-rank dynamics, where techniques such as compressed 
sensing or $\ell_0$-norm optimization may be employed for state reconstruction. 
Furthermore, the proposed approach can be generalized to multi-agent or 
networked tank systems, where distributed estimation and decentralized control 
strategies are required to handle system coupling, communication constraints, 
and scalability.

\bibliographystyle{ieeetr}  
\bibliography{reference}

\end{document}